\documentstyle[11pt,newpasp,twoside,epsf]{article}
\def\edcomment#1{\iffalse\marginpar{\raggedright\sl#1\/}\else\relax\fi}
\reversemarginpar

\begin{document}
\title{Physical Conditions in Lyman Break Galaxies Derived From Rest-Frame UV Spectra}
\author{Alice E. Shapley and Charles C. Steidel}
\affil{Palomar Observatory, Caltech 105-24, Pasadena, CA 91125, USA}
\author{Kurt L. Adelberger}
\affil{Harvard-Smithsonian Center for Astrophysics, 60 Garden St., Cambridge, MA 02138, USA}
\author{Max Pettini}
\affil{Institute of Astronomy, Madingley Road, Cambridge, CB3 0HA, UK}

\begin{abstract}
We present the results of detailed studies of the astrophysical conditions
in $z\sim 3$ Lyman Break Galaxies (LBGs), placing particular
emphasis on what is learned from LBG rest frame UV spectra.
By drawing from our database
of $\sim 1000$ spectra, and constructing higher S/N composite spectra from
galaxies grouped according to properties such as luminosity, extinction,
morphology, and environment, we can show how the rest-frame UV
spectroscopic properties systematically depend on other galaxy properties.
Such information is crucial to understanding the detailed nature of LBGs,
and their impact on the surrounding IGM.
\end{abstract}
\section{Introduction}
Until now, Lyman Break Galaxy (LBG) rest-frame
UV spectra have been primarily used to measure
redshifts. The measured redshifts confirm the 
high-redshift nature of galaxy candidates selected
by their distinctive broadband optical colors (Steidel et al. 1996);
enable the study of the spatial clustering of these
galaxies (Adelberger et al. 1998); and, when combined with the apparent
magnitudes and colors of LBGs, can be used to
construct the rest-frame UV luminosity function
and UV luminosity density at $z\sim 3$ (Steidel et al. 1999).

\section{LBG Spectra}
\subsection{A Special Case: MS1512-cB58}
One notable exception is the galaxy MS1512-cB58, a
LBG at $z=2.73$, discovered serendipitously by Yee et al. (1996)
during the CNOC cluster redshift survey. cB58 has an apparent $V$ magnitude
of 20.6, due to lensing by a foreground cluster at $z=0.37$,
which boosts its apparent luminosity by about a factor of $\sim 30$
(Seitz et al. 1998). Due to the incredibly bright
nature of cB58, a wealth of information has been extracted
from medium-resolution ($R\simeq 5000$) studies of its rest-frame UV spectrum
(Pettini et al. 2000, 2001). The velocity profiles of low and high-ionization
interstellar metal absorption features have been characterized in detail;
the weakest interstellar metal transitions have been used together with 
the damped Lyman-$\alpha$ absorption profile to determine the
abundance pattern in cB58 (an $\alpha$/Fe enhancement
indicative of a young stellar population, and an abundance of 
$\sim 2/5 Z_{\odot}$ for the $\alpha$ elements); CIV and SIV P-Cygni 
stellar wind profiles have been used as independent probes of the stellar
population and metallicity of cB58; weak stellar
absorption features have been used to precisely measure the
systemic velocity of the stars in cB58, relative to which the
redshifts of Lyman $\alpha$ emission and interstellar
absorption indicate offsets of several hundred km s$^{-1}$; 
finally, the strengths of the strongest interstellar absorption features
(which have zero transmission at line center) have been used
to infer a high covering fraction of absorbing material, through
which negligible Lyman continuum emission can escape 
(Heckman et al. 2001).

\subsection{Typical LBG Spectra}
In stark contrast to the spectra of cB58, the $12.5 \: {\rm \AA}$ resolution
spectra obtained for unlensed LBGs 
($R_{AB}=23-25.5$) after 1.5 hours of integration 
with the Keck telescopes' Low Resolution Imaging Spectrometer (Oke et al. 1995)
typically have signal-to-noise ratios
of only a few, and therefore do not allow for the same type of
detailed study on an individual basis. The only features visible
in even the best spectra are Lyman $\alpha$ (in emission,
absorption, or both), and the strongest interstellar absorption
features, whose saturated equivalent widths are indicative of some combination
of velocity widths and covering factor, but not abundance measures.

\subsection{Composite LBG spectra}
While individual LBG spectra provide limited information,
our group has assembled a database of almost 1000 spectra of $z \geq 2$
galaxies over the past five years. Subsamples of 
galaxies can be drawn from the database with specific
criteria, and used to construct higher S/N composite spectra.
For example, a composite spectrum constructed from 29 LBGs at 
$\langle z \rangle = 3.4 \pm 0.09$ indicates significant 
positive flux bluewards of the Lyman cutoff at
$912 \; {\rm \AA}$ (Steidel, Pettini, \& Adelberger 2001). 
If this composite spectrum is taken to be
representative of LBGs at $z\sim 3$, then the LBG contribution to the
ionizing background at $z\sim3$ could exceed that of QSOs at similar 
redshifts by as much as a factor of 5. It should be noted
that this composite spectrum inhabits an extreme of LBG UV spectral
parameter space, with strong Lyman $\alpha$ emission,
a blue UV continuum slope, and interstellar absorption lines
whose equivalent widths are roughly half the strength of those seen
in cB58. Also, separate composite spectra constructed for 
16 ``young'' ($t\leq $ 35 Myr) and 16 ``old'' ($t\sim $ 1 Gyr) LBGs, 
(whose stellar population parameters were determined from optical/IR photometry)
exhibited systematic differences. The ``old'' spectrum had much stronger
Lyman $\alpha$ emission, while the ``young'' spectrum had
stronger interstellar absorption lines (indicating larger
velocity widths or covering fraction of gas), and more pronounced
P-Cygni high-ionization stellar wind features (Shapley et al. 2001).

\section{Preliminary Results}
Building on the above pilot studies which included composite
LBG rest-frame UV spectra, we are now undertaking a systematic
analysis of our entire database of spectra. These spectra
primarily probe outflowing material which is being
propelled by the mechanical energy input from active
star-formation. In order to study the physical parameters
of these outflows, we will determine how line strengths 
and kinematics vary in composite
spectra constructed along sequences of properties such as 
UV-luminosity and color, bolometric
luminosity, morphology (from HST), redshift, and Lyman $\alpha$ profile. 

\subsection{Lyman $\alpha$ vs. Reddening, Equivalent Widths, and Kinematics}
As a first step, we have divided our sample according 
to a crude classification of Lyman $\alpha$ profile, and see intriguing
variations in the resulting composite spectra. Based on the result of 
Steidel et al. 2000 that the distribution of LBG rest-frame 
Lyman $\alpha$ equivalent widths has a median of 
$0 \: {\rm \AA}$ and ranges from $-100 \: {\rm \AA}$
to $100 \: {\rm \AA}$, we divide galaxies into Lyman $\alpha$ ``emission'',
``emission/absorption'', and ``absorption'' subsamples. Figure 1
shows the resulting composite spectra, each of which contains more
than 200 galaxies, and has a S/N per pixel of $\sim 25$. 

We find that the UV continuum shape becomes systematically redder with
increasing Lyman $\alpha$ absorption strength, indicating
that both the UV continuum and Lyman $\alpha$ photons are affected
by dust extinction, and providing information about the dust-gas 
geometry in LBGs. We also find that the equivalent widths of 
low-ionization interstellar features of SiII, OI, CII, FeII, and AlII,
associated with the neutral ISM, increase with larger Lyman
$\alpha$ absorption strength. As stated above, these equivalent
widths are not indicative of abundances, but rather velocity
widths, and possibly ISM covering factor. Therefore, the larger
equivalent widths associated with the ``absorption'' spectrum
indicate a larger velocity field, or possibly an ISM which
is more opaque to ionizing radiation. 
We note that the spectrum of cB58 most closely
resembles that of the composite ``absorption'' spectrum, whereas the
composite spectrum used to measure the LBG Lyman continuum leakage
has Lyman $\alpha$ emission and metal absorption strengths which
are similar to that of the ``emission'' composite spectrum. Finally, 
kinematic properties of LBG outflows seem to depend on the Lyman $\alpha$
profile. The mean rest-frame velocity offset between Lyman $\alpha$ emission
and interstellar absorption lines is $\Delta v= 630 \; {\rm km}^{-1}{\rm s}$ 
($\Delta z = 0.008$), however, the velocity offset decreases with
increasing Lyman $\alpha$ emission strength. We have yet to determine
whether this trend is dominated by ISM geometry, or evolutionary effects.

\begin{figure}[t]
\plotfiddle{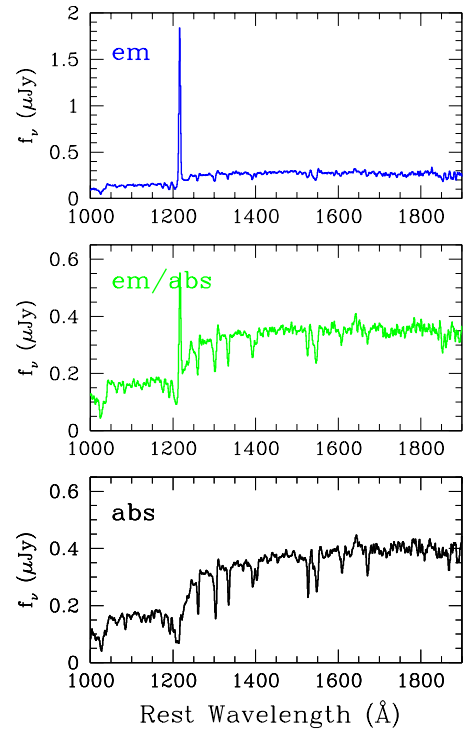}{3.5in}{0}{50}{50}{-150}{-75}
\caption{Composite spectra constructed according to Lyman $\alpha$ profile. Each composite spectrum contains more than 200 galaxies, and has a S/N per pixel of $\sim 25$. The interstellar absorption line strengths increase and the continuum becomes redder with increasing Lyman $\alpha$ absorption strength.}
\end{figure}

\section{Conclusions and Future Work}
Rapid star-formation has a profound
effect on the ISM of LBGs, as seen in the large absorption
equivalent widths and significant
velocity offsets between rest-frame UV emission and absorption
features. The effects of star-formation also extend to the 
surrounding inter-galactic medium, through the leakage
of Lyman continuum emission, and from shock heating by
outflowing material. A systematic analysis of the 
rest-frame UV spectroscopic of LBGs will help us understand
how star-formation transforms both galaxies and
their surrounding environment at $z\sim 3$.


\begin{references}

Adelberger, K.L.  et al. 1998, \apj, 505, 18A

Heckman, T.M. et al. 2001, \apj, 558, 56H

Oke, J.B. et al. 1995, \pasp, 107, 375O

Pettini M. et al. 2000, \apj, 528, 96P

Pettini M. et al. 2001, \apj ~accepted, (astro-ph/0110637)

Seitz S. et al. 1998, \mnras, 298, 945S

Shapley A.E. et al. 2001, \apj, 562, 95S

Steidel C.C. et al. 1996, \apj, 462, L17S

Steidel C.C. et al. 1999, \apj, 519, 1S

Steidel C.C. et al. 2000, \apj, 532, 170S

Steidel, C.C., Pettini, M., \& Adelberger, K.L. 2001, \apj, 546, 665S

Yee H.K.C. et al. 1996, \aj, 111, 1783Y
\end{references}
\end{document}